\documentclass[12pt,a4]{article}
\newcommand{\be}{\begin{equation}}
\newcommand{\ee}{\end{equation}}
\newcommand{\bi}[1]{\vspace{-3mm} \bibitem{#1}}
\begin{document}
{\large \it CHAOS, \ Vol.14. No.1. (2004) pp.123-127.}
\vskip 11mm
\begin{center}
{\Large \bf Fractional Generalization of Liouville Equations}
\vskip 5 mm
{\Large \bf Vasily E. Tarasov } \\

\vskip 3mm
{\it Skobeltsyn Institute of Nuclear Physics,
Moscow State University, Moscow 119992, Russia}

{E-mail: tarasov@theory.sinp.msu.ru}
\end{center}
\begin{abstract}
In this paper fractional generalization of Liouville equation
is considered.
We derive fractional analog of normalization condition
for distribution function.
Fractional generalization of the Liouvile equation
for dissipative and Hamiltonian systems was derived
from the fractional normalization condition.
This condition is considered considered as a normalization 
condition for systems in fractional phase space.
The interpretation of the fractional space is discussed.
\end{abstract}

\vskip 3mm
PACS {05.20.-y, 05.45.-a}


\vskip 7mm
{\bf
We call a fractional equation a differential equation
that uses fractional derivatives or integrals.
Fractional derivatives and integrals have found many
applications in recent studies of scaling phenomena.
We formulate fractional analog of main
integro-differential equation to describe
some scaling process - Liouville equation.
Usually used for scaling phenomena Fokker-Planck
equation can be derived from Liouville equation.
Therefore it is interesting to consider a fractional
generalization of the Liouville equation.
To derive fractional equation for a distribution
function we must consider a fractional analog of
the normalization condition for distribution function.
Most of the fractional equations for distribution function
does not use correspondent normalization condition.
Therefore these equations (with fractional coordinate
derivatives) can be incorrect equations.
In this paper fractional Liouvile equation for dissipative
systems is derived from the normalization condition.
The coordinate fractional integration for this
normalization condition is used. }

\newpage
\section{Introduction}

Fractional derivatives and integrals have found many
applications in recent studies of scaling phenomena \cite{1,2,3,4,Zas2}.
The main aim of most of these papers is to formulate fractional
integro-differential equations to describe some scaling process.
Modifications of equations governing physical processes such as
the Langevin equation \cite{L}, diffusion equations, 
and Fokker-Plank equation  have been suggested
\cite{5}-\cite{10} which incorporate fractional
derivatives with respect to time.
It was shown in Ref. \cite{Zas} that the chaotic Hamiltonian
dynamics of particles can be described by using fractional
generalization of the Fokker-Planck-Kolmogorov equation.
In Ref. \cite{Zas}, coordinate fractional derivatives
in the Fokker-Planck equation were used.

It is known that Fokker-Planck equation can be derived
from Liouville equation \cite{Is,RL}.
Therefore it is interesting to consider a fractional generalization
of Liouville equation and Bogoliubov hierarchy equations.
We call a fractional equation a differential equation
that uses fractional derivatives or integrals.

To derive fractional equations for a distribution
function we must consider a fractional analog of
the normalization condition for distribution function.
Fractional Liouville equation for dissipative systems
is derived from the normalization condition.
In this paper, the coordinate fractional integration for
normalization condition is used.
This condition is considered considered as a normalization 
condition for systems in fractional phase space.
If any fractional equation for
distribution function does not use correspondent
normalization condition, then this equation
(with fractional coordinate derivatives) can be incorrect.

In Sec. 2 the normalization condition for
distribution function and notations
are considered.
In Sec. 3 we derive the Liouvile equation
from the normalization condition.
In Sec. 4 the physical interpretation of fractional
normalization condition is considered.
Finally, a short conclusion is given in Sec. 5.

\section{Normalization condition}

Let us consider a distribution function $\rho(x,t)$ for
$x\in R^{1}$. Let $\rho(x,t) \in L_{1}(R^{1})$, where
$t$ is a parameter. Normalization condition has the form
\[ \int^{+\infty}_{-\infty} \rho(x,t) dx =1. \]
This condition can be rewritten in the form
\be \label{II} \int^{y}_{-\infty} \rho(x,t) dx+
\int^{+\infty}_{y} \rho(x,t) dx=1,\ee
where  $y\in (-\infty,+\infty)$.

Let $\rho(x,t) \in L_{p}(R^{1})$, where $1<p<1/\alpha$.
Fractional integrations on $(-\infty,y)$ and $(y,+\infty)$
are defined \cite{SKM} by
\be \label{I+} (I^{\alpha}_{+}\rho)(y,t)=\frac{1}{\Gamma(\alpha)}
\int^{y}_{-\infty} \frac{\rho(x,t)dx}{(y-x)^{1-\alpha}} , \ee
\be \label{I-} (I^{\alpha}_{-}\rho)(y,t)=\frac{1}{\Gamma(\alpha)}
\int^{+\infty}_{y} \frac{\rho(x,t)dx}{(x-y)^{1-\alpha}} . \ee
Using these notations, Eq. (\ref{II}) has the form
\[ (I^{1}_{+}\rho)(y,t)+(I^{1}_{-}\rho)(y,t)=1. \]
Using definitions (\ref{I+}) and (\ref{I-}) we can get the
fractional analog of normalization condition (\ref{II}), 
\[ (I^{\alpha}_{+}\rho)(y,t)+(I^{\alpha}_{-}\rho)(y,t)=1. \]
Equations (\ref{I+}) and (\ref{I-}) can be rewritten
in the form
\be \label{Ipm} (I^{\alpha}_{\pm}\rho)(y,t)=
\frac{1}{\Gamma(\alpha)}
\int^{\infty}_{0} x^{\alpha-1}\rho(y\mp x,t )dx . \ee
This leads to the normalization condition
\be \label{nc2} \frac{1}{\Gamma(\alpha)}
\int^{\infty}_{0} x^{\alpha-1} [\rho(y-x,t)+\rho(y+x,t) ] dx=1 . \ee
If we denote
\be \label{til} \tilde \rho(x,t)=\rho(y-x,t)+\rho(y+x,t) \ee
and
\be d\mu_{\alpha} (x)=\frac{x^{\alpha-1}}{\Gamma(\alpha)} dx,  \ee
then
condition (\ref{II}) has the form
\be \label{nc3} \int^{\infty}_{0} \tilde \rho (x,t) d \mu_{\alpha} (x)=1. \ee

Note that substituting $y=ct$ in (\ref{til}), we get the sum
\[ \tilde \rho(x_{t},t)=\rho(ct-x_{t},t)+\rho(ct+x_{t},t) . \]
This sum can be considered as a sum of right and back waves of
the distribution functions.

\section{Liouville equation}

Let us consider a domain $B_{0}$ for the time $t=0$.
In the Hamilton picture we have
\[ \int_{B_{t}} \tilde \rho(x_{t},t) d\mu_{\alpha} (x_{t})=
\int_{B_{0}}\tilde \rho(x_{0},0) d\mu_{\alpha} (x_{0}). \]
Using the replacement of variables $x_{t}=x_{t}(x_{0})$,
where $x_{0}$ is a Lagrangian variable, we get
\[ \int_{B_{0}} \tilde \rho(x_{t},t) x^{\alpha-1}_{t}
\frac{\partial x_{t}}{\partial x_{0}} dx_{0}=
\int_{B_{0}} \tilde \rho(x_{0},0) x^{\alpha-1}_{0} dx_{0}. \]
Since $B_{0}$ is an arbitrary domain we have
\[ \tilde \rho (x_{t},t) d\mu_{\alpha} (x_{t})=
\tilde \rho (x_{0},0) d\mu_{\alpha} (x_{0}), \]
or
\[ \tilde \rho(x_{t},t) x^{\alpha-1}_{t}
\frac{\partial x_{t}}{\partial x_{0}} =
\tilde \rho(x_{0},0) x^{\alpha-1}_{0} . \]
Differentiating this equation in time $t$, we obtain
\[ \frac{d \tilde \rho(x_{t},t)}{dt} x^{\alpha-1}_{t}
\frac{\partial x_{t}}{\partial x_{0}}
+ \tilde \rho(x_{t},t) \frac{d}{dt} \Bigl(x^{\alpha-1}_{t}
\frac{\partial x_{t}}{\partial x_{0}}\Bigr)=0 , \]
or
\be \label{Liu} \frac{d \tilde \rho(x_{t},t)}{dt}
+ \Omega_{\alpha}(x_{t},t) \tilde \rho(x_{t},t)=0 , \ee
where $d/dt$ is a total time derivative
\[ \frac{d}{dt}=\frac{\partial}{\partial t}+F_{t}
\frac{\partial}{\partial x_{t}} . \]
The function
\[ \Omega_{\alpha}(x_{t},t)=
\frac{d}{dt} ln \Bigl(x^{\alpha-1}_{t}
\frac{\partial x_{t}}{\partial x_{0}}\Bigr) \]
describes the velocity of (phase space) volume change.
Equation (\ref{Liu}) is a fractional Liouville equation
in the Hamilton picture.
If the equation of motion has the form
\[ \frac{dx_{t}}{dt}=F_{t}(x), \]
then the function $\Omega_{\alpha}$ is defined by
\[ \Omega_{\alpha}(x_{t},t)=
\frac{d}{dt} \Bigl( ln \ x^{\alpha-1}_{t} +
ln \ \frac{\partial x_{t}}{\partial x_{0}}\Bigr)=
(\alpha-1) \frac{1}{x_{t}} \frac{dx_{t}}{dt} +
\frac{\partial}{\partial x_{t}} \frac{dx_{t}}{dt} . \]
As the result we have
\[ \Omega_{\alpha}(x_{t},t) = \frac{(\alpha-1)F_{t}}{x_{t}} +
\frac{\partial F_{t}}{\partial x_{t}}. \]

The normalization in the phase space is derived by analogy
with a normalization in the configuration space.
The fractional normalization  condition in the phase space
\be \label{fnc2} \int^{\infty}_{0}
\int^{\infty}_{0} \tilde \rho (q,p,t) d \mu_{\alpha} (q,p)=1, \ee
where $d\mu_{\alpha} (q,p)$ has the form
\be \label{mu0} d\mu_{\alpha} (q,p)=
d\mu_{\alpha} (q) \wedge d\mu_{\alpha} (q)=
\frac{dq^{\alpha} \wedge dp^{\alpha}}{(\alpha \Gamma(\alpha))^{2}}
=\frac{(qp)^{\alpha-1} }{\Gamma^2(\alpha)} dq \wedge dp.  \ee
The distribution function $\tilde \rho(q,p,t)$
in the phase space is defined by
\[ \tilde \rho(q,p,t)=\rho(q'-q,p'-p,t)+\rho(q'+q,p'-p,t)+
\rho(q'-q,p'+p,t)+\rho(q'+q,p'+p,t) . \]

Let us use the well known transformation
\be \label{qpt0} dq_t \wedge d p_t
= \{q_t,p_t\}_0 dq_0 \wedge d p_0 ,\ee
where $\{q_t,p_t\}_0$ is Jacobian which
is defined by the determinant
$$\{q_t,p_t\}_0 =det  \frac{\partial(q_t,p_t)}{\partial (q_0,p_0)}
=det \ \left( \begin{array}{cc}
{\partial q_{kt}}/{\partial q_{l0}}&
{\partial q_{kt}}/{\partial p_{l0}}\\
{\partial p_{kt}}/{\partial q_{l0}}&
{\partial p_{kt}}/{\partial p_{l0}}
\end{array}
\right).$$

Using $\tilde \rho_t d\mu_{\alpha} (q_t,p_t)
=\tilde \rho_0 d\mu_{\alpha} (q_0,p_0)$, we get the relation
\be \label{mu3}
\tilde \rho_t \frac{(q_tp_t)^{\alpha-1}}{ \Gamma^2(\alpha)}
dq_t \wedge d p_t =\tilde \rho_0
\frac{(q_0p_0)^{\alpha-1}}{ \Gamma^2(\alpha)}
dq_0 \wedge d p_0  . \ee
Using (\ref{qpt0}), we have condition (\ref{mu3}) in the form
\be \label{nn1} \tilde \rho_t(q_tp_t)^{\alpha-1}\{q_t,p_t\}_0 =
(q_0p_0)^{\alpha-1} \tilde \rho_0  \ee
Let us write condition (\ref{nn1}) in more simple form
\be \label{nn2} \tilde \rho_t \{q^{\alpha}_t,p^{\alpha}_t\}_0 =
\alpha^2 (q_0p_0)^{\alpha-1} \tilde \rho_0 . \ee
The time derivatives of this equation
lead to the fractional Liouville equation
\be \label{Liu3} \frac{d \tilde \rho(q_{t},p_t,t)}{dt}
+\Omega_{\alpha}(q_{t},p_t,t) \tilde \rho(q_{t},p_t,t)=0 , \ee
where the function $\Omega_{\alpha}$ is defined by
\be \label{02}
\Omega_{\alpha}(q_{t},p_t,t)= \{q^{\alpha}_t,p^{\alpha}_t\}^{-1}_0
\frac{d}{dt}\{q^{\alpha}_t,p^{\alpha}_t\}_0=
\frac{d}{dt} ln \{q^{\alpha}_t,p^{\alpha}_t\}_0 . \ee
In the usual notations we have
\be \label{o3} \Omega_{\alpha}(q_{t},p_t,t)=
\frac{d}{dt} ln \ det
\frac{\partial(q^{\alpha}_t,p^{\alpha}_t)}{\partial (q_0,p_0)} . \ee
Using well-known relation $ln\ det \ A=Sp \ ln \ A$,
we can write the $\alpha$-omega function in the form
\[ \Omega_{\alpha}=\{\frac{dq^{\alpha}_t}{dt},p^{\alpha}_t\}_{\alpha}+
\{q^{\alpha}_t,\frac{dp^{\alpha}_t}{dt}\}_{\alpha} , \]
where
\[ \{A,B\}_{\alpha}=\frac{\partial A}{\partial q^{\alpha}}
\frac{\partial B}{\partial p^{\alpha}}-
\frac{\partial A}{\partial p^{\alpha}}
\frac{\partial B}{\partial q^{\alpha}}  . \]
In the general case ($\alpha \not=1$) the function
$\Omega_{\alpha}$ is not equal to zero ($\Omega_{\alpha} \not=0$)
for Hamiltonian systems.
If $\alpha=1$, we have $\Omega_{\alpha} \not=0$ only
for non-Hamiltonian systems.

It is easy to see that any system which is defined by the equations
\be \frac{dq^{\alpha}_t}{dt}=\frac{p_t}{m},
\quad \frac{dp^{\alpha}_t}{dt}=f(q_t), \ee
has the $\alpha$-omega function equal to zero
$\Omega_{\alpha}=0$.
This system can be called a fractional
nondissipative system.
For example, a fractional oscillator is defined by
the equation
\be \frac{dq^{\alpha}_t}{dt}=\frac{p^{\alpha}_t}{m},
\quad \frac{dp^{\alpha}_t}{dt}=-m\omega^2 q^{\alpha}_t . \ee

The $\alpha$-omega function can be rewritten in the form
\be \Omega_{\alpha}(q_{t},p_t,t)=
(\alpha-1)\Bigl(q^{-1}_{t}\frac{dq_t}{dt}+
p^{-1}_{t}\frac{dp_t}{dt}\Bigr)+
\{\frac{dq_t}{dt},p_t\}_1+\{q_t,\frac{dp_t}{dt}\}_1 , \ee
where $\{.,.\}_1$ is usual Poisson bracket.
If the Hamilton equations have the form
\be \label{H3}
\frac{dq_t}{dt}=G(q_t,p_t), \quad \frac{dp_t}{dt}=F(q_t,p_t), \ee
then the $\alpha$-omega function is defined by
\be \label{om3}  \Omega_{\alpha}(q,p)=
(\alpha-1)\Bigl(q^{-1} G(q,p)+p^{-1} F(q,p)\Bigr)+
\{G,p\}_1+\{q,F\}_1. \ee
This relation allows to derive $\Omega_{\alpha}$
for all dynamical systems (\ref{H3}).
It is easy to see that the usual nondissipative system
\be \label{eq1} \frac{dq_t}{dt}=\frac{p_t}{m},
\quad \frac{dp_t}{dt}=f(q_t), \ee
has the $\alpha$-omega function
\[ \Omega_{\alpha}(q,p)=(\alpha-1)(mqp)^{-1}(p^2+mq f(q)) \]
and can be called a fractional dissipative system.
For example, the linear harmonic oscillator ($f(q)=-m\omega^2 q$)
\[ \Omega_{\alpha}(q,p)=
(\alpha-1)(mqp)^{-1}(p^2-m^2\omega^2 q^2) , \]
is a fractional dissipative system.

\section{Interpretation}

The fractional normalization condition can be considered as a
normalization condition for the distribution function in a
fractional phase space.
In order to use this interpretation we must define
a fractional phase space.

The first interpretation of the fractional phase space
is connected with fractional dimension.
This interpretation follows from
the well-known formulas for dimensional regularizations \cite{Col}:
\be \label{dr} \int \rho(x) d^{n} x =
\frac{2 \pi^{n/2}}{\Gamma(n/2)}
\int^{\infty}_{0} \rho(x)  x^{n-1} dx  . \ee
Using Eq. (\ref{dr}), we get
that the fractional normalization condition (\ref{nc3})
can be considered as a normalization condition
in the fractional dimension space
\be \label{fnc-2} \frac{\Gamma(\alpha/2) }{2 \pi^{\alpha/2} \Gamma(\alpha)}
\int \tilde \rho (x,t) d^{\alpha} x=1  \ee
up to the numerical factor
$\Gamma(\alpha/2) /( 2 \pi^{\alpha/2} \Gamma(\alpha))$.

The second interpretation is connected with the fractional measure
of phase volume.
The parameter $\alpha$  defines the space with
the fractional phase volume
\[ V_{\alpha}=\int_B d \mu_{\alpha}(q,p) . \]
It is easy to prove that the velocity of the fractional 
phase volume change is defined by 
\[ \frac{dV_{\alpha}}{dt}=
\int_B \Omega_{\alpha}(q,p,t) d\mu_{\alpha}(q,p) .\]
Note that the volume element of fractional phase space can be realized 
by fractional exterior derivatives \cite{CN}
\be d^{\alpha}=\sum^{n}_{k=1}dq^{\alpha}_k \frac{\partial^{\alpha}}{(\partial 
(q_k-a_k))^{\alpha}}+\sum^{n}_{k=1}
dp^{\alpha}_k \frac{\partial^{\alpha}}{(\partial (p_k-b_k))^{\alpha}}, \ee
in the form
\be dq^{\alpha} \wedge dp^{\alpha}= \Bigl(\frac{4}{\Gamma^2(2-\alpha)}+
\frac{1}{\Gamma^2(1-\alpha)}\Bigr)^{-1} 
(qp)^{\alpha-1} d^{\alpha}q \wedge d^{\alpha}p . \ee

The system can be called a fractional dissipative system
if a fractional phase volume changes, i.e., $\Omega_{\alpha} \not=0$.
The system which is a nondissipative system
in the usual phase space,
can be a dissipative system in the fractional phase space.
The fractional analog of the usual conservative
Hamiltonian nondissipative system is defined by the equations
\be \label{fhs} \frac{dq^{\alpha}_{k}}{dt}=\frac{g_{k}(p)}{m},\quad
 \frac{dp^{\alpha}_{k}}{dt}=f_k(q). \ee
The usual nondissipative systems (\ref{eq1})
are dissipative in the fractional phase space.

In the general case, the fractional system is a system 
in the fractional phase space.
We shall say that a system is called a fractional system if this system
can be described by the fractional powers of coordinates and momenta,
\[ q^{\alpha}_k=|q_k|^{\alpha}, \quad p^{\alpha}_k=|p_k|^{\alpha} , \]
where $k=1,...,n$.

The fractional systems allow to consider the interpretation
of the fractional normalization condititon which is used to
derive the fractional Liouville equation.
The fractional normalization condition for the distribution function
can be considered as a normalization condition
for the systems in the fractional phase space.

The Hamilton equations for the fractional system have the form
\be \label{HE} \frac{dq^{\alpha}_{k}}{dt}=\frac{p^{\alpha}_{k}}{m}, \quad
\frac{dp^{\alpha}_{k}}{dt}=F_{k}(q^{\alpha},p^{\alpha}) . \ee
Obviously, that the equation $dp_{k}/dt=F_{k}$ can be rewritten
in the fractional form. Multiplying both sides of this equation
by $\alpha p^{\alpha-1}$, we obtain
$dp^{\alpha}_{k}/dt=\alpha p^{\alpha-1}F_{k}$.
However the equation $dq_{k}/dt=p_{k}/m$ cannot
be rewritten in the fractional form
$dq^{\alpha}_{k}/dt=p^{\alpha}_{k}/m$.


The fractional conservative Hamiltonian system is described
by the equation
\be \frac{dq^{\alpha}_{k}}{dt}=
\frac{\partial H(q^{\alpha},p^{\alpha})}{\partial p^{\alpha}_{k}},
\quad \frac{dp^{\alpha}_{k}}{dt}=
-\frac{\partial H(q^{\alpha},p^{\alpha})}{\partial q^{\alpha}_{k}}, \ee
or
\be \frac{dq^{\alpha}_{k}}{dt}=\{q^{\alpha}_{k},H\}_{\alpha},
\quad \frac{dp^{\alpha}_{k}}{dt}=\{p^{\alpha}_{k},H\}_{\alpha}. \ee
Here $H=H(q^{\alpha},p^{\alpha})$ is a fractional analog of the
Hamiltonian function
\be \label{Ha} H(q^{\alpha},p^{\alpha})=
\sum^n_{k=1} \frac{p^{2\alpha}_{k}}{2m} +U(q^{\alpha}) . \ee
The fractional system can be considered as a nonlinear system with
\be \label{Hqp} H(q^{\alpha},p^{\alpha})=
\sum^n_{k=1}\frac{1}{2} g_{kl}(q,p) p_k p_l +U(q) .\ee
Note that the Hamiltonian (\ref{Hqp}) defines a
nonlinear one-dimensional sigma-model \cite{Tar-mpla,Tar-pl}
with metric
\[ g_{kl}(q,p)= m^{-1} p^{2(\alpha -1)}_k \delta_{kl} .  \]

It is easy to see that fractional systems (\ref{Ha})
can lead to the non-Gaussian statistics.
The interest in and relevance of fractional kinetic equations
is a natural consequence of the realization of the importance of
non-Gaussian statistics of many dynamical systems. There is already
a substantial literature studying such equations in one or more
space dimensions.

Note that the classical (nonlinear) dissipative systems can have
canonical Gibbs distribution as a solution of stationary
Liouville equations for this dissipative system \cite{Tar-mplb}.
Using the methods \cite{Tar-mplb}, it is easy to prove
that some of fractional dissipative systems can have
fractional canonical Gibbs distribution
\[ \rho(q,p)=Z(T)exp -\frac{H(q^{\alpha},p^{\alpha})}{kT} , \]
as a solution of the fractional Liouville equations
\be \label{LE} \frac{\partial \rho}{\partial t}+
\frac{p^{\alpha}_{k}}{m}\frac{\partial \rho}{\partial q^{\alpha}_{k}}+
\frac{\partial}{\partial p^{\alpha}_{k}}
\Bigl(F_{k}(q^{\alpha},p^{\alpha}) \rho\Bigr)=0 . \ee
Here the function $H(q^{\alpha},p^{\alpha})$ is defined by (\ref{Ha}).

\section{Conclusion}

Derivatives and integrals of fractional order have found many
applications in studies of scaling phenomena \cite{1,2,3,4,Zas2}.
In this paper we formulate fractional analog of main
integro-differential equation to describe some scaling process -
Liouville equation.
We consider the fractional analog of the normalization condition
for the distribution function.
Fractional Liouvile equation for dissipative systems
is derived from the normalization condition.
In this paper, the coordinate fractional integration
for the normalization condition is used.
This fractional normalization condition can be considered
as a simulating unconventional environment for systems
with fractional dimensional phase space or phase space with
fractional powers of coordinates and momenta.
Note that the adoption of fractal formalism yields properties 
that the ordinary formalism would produce only in 
the case where the system is made non-Hamiltonian by the presence of 
an environment, whose influence can be mimicked by means of friction, 
for instance.

Suggested fractional Liouville equation allows to formulate
the fractional equation for quantum dissipative systems \cite{Tarkn1}
by methods suggested in Refs. \cite{Tarpla1,Tarmsu}.
In general, we can consider this dissipative quantum systems
as quantum computer with mixed states \cite{Tarjpa}.
These dissipative quantum systems can have stationary states
\cite{Tarpla2}.
Stationary states of dissipative quantum systems
can coincide with stationary states of Hamiltonian
systems \cite{Tarpre}. \\

This work was partially supported by the RFBR Grant No. 02-02-16444




\end{document}